\documentclass[10pt,twocolumn]{article}

\usepackage[margin=0.75in]{geometry}
\usepackage[T1]{fontenc}
\usepackage{lmodern}
\usepackage{microtype}
\usepackage{booktabs}
\usepackage{xcolor}
\usepackage{graphicx}
\usepackage{caption}
\usepackage{subcaption}
\usepackage{tikz}
\usepackage{pgfplots}
\usepackage{pgfplotstable}
\pgfplotsset{compat=1.18}
\usepackage[hidelinks]{hyperref}

\IfFileExists{data/palette.tex}{\definecolor{vcanthropic}{RGB}{198,93,42}
\definecolor{vcdeepseek}{RGB}{150,80,190}
\definecolor{vcgoogle}{RGB}{58,120,216}
\definecolor{vcopenai}{RGB}{16,140,110}
}{}

\IfFileExists{data/macros.tex}{

\newcommand{\dataVendorsNice}{Anthropic, DeepSeek, Google, OpenAI}

\newcommand{\dataSizes}{40k, 100k}
\newcommand{\dataIdles}{0, 60, 300, 600}

\newcommand{\keepaliveMinHit}{98}
\newcommand{\bestSavingRatio}{\ensuremath{12.5\times}}

}{}
\providecommand{\dataSizes}{--}%
\providecommand{\dataIdles}{--}%
\providecommand{\keepaliveMinHit}{--}\providecommand{\bestSavingRatio}{--}%
\providecommand{\dataVendorsNice}{--}

\title{\vspace{-2em}Keeping the Cache Warm Pays:\\Keepalive Economics for Agentic Workloads}
\author{Maxim Khailo \\ \small blog.mempko.com \\ \small \texttt{max@mempko.com}}
\date{\today}

\begin{document}
\maketitle

\begin{abstract}
Frontier LLM providers cache a prompt's processed prefix so that a follow-up
request sharing it pays ${\sim}10\%$ of the input price and skips most of the
prefill latency. Agentic workloads systematically destroy this benefit: the
agent sends a request, runs a tool or waits for approval for minutes, and by
the time the follow-up is sent the cached prefix has been evicted, so the agent
pays the full prefill again. A client-side \emph{keepalive}, replaying the
prefix on a timer during the pause, prevents this, and it is \emph{individually
rational}: across Anthropic, OpenAI, Google, and DeepSeek we show that a
keepalive holds the prefix warm through gaps where idle baselines are evicted,
cutting the post-pause request cost by up to \bestSavingRatio{}. The strategic
question is the ping \emph{frequency}, and it has a clean answer: keepalive
cost falls monotonically in the interval, so the economical choice is the
largest interval safely under the provider's TTL, about 4 minutes at
Anthropic's 5-minute TTL rather than the 30-second convention. The keepalive
saves inside a \emph{paying band} bounded below by the provider's measured
eviction point and above by $I_{\max}\approx\tau(w/r-1)$. We locate each
eviction point (Anthropic ${\sim}5$, DeepSeek ${\sim}10$, OpenAI ${\sim}30$
minutes; Google never converges) and confirm at a 30-minute pause that the
keepalive saves on Anthropic \emph{and} OpenAI, while DeepSeek's re-prefill is
too cheap to insure and Google never reliably evicts, leaving those two a
latency win only. Because the benefit is real and bounded only by each user's own bill, rational
adoption is universal adoption; and since cache residency is priced per read
rather than per token-hour, a keepalive-saturated tier gives LRU eviction
nothing to rank. We argue this externality will push providers to meter cache
residency directly, and one already does. We derive the operator's policy until
then.
\end{abstract}

\section{Introduction}
Modern LLM inference reuses the attention KV cache of a shared prompt
\emph{prefix}: a request beginning with recently processed bytes reads cached
state instead of recomputing
it~\cite{pope2023efficiently,kwon2023efficient,gim2024promptcache}. Exposed as
``prompt caching''~\cite{anthropic-caching,openai-caching,google-caching}, this
makes cached input tokens roughly $10\times$ cheaper and removes most of the
prefill latency.

Agentic workloads are the worst-case user of this mechanism. An agent's loop
is: think, act, wait. It sends a request, then runs a tool (a build, a test
suite, a deployment, a wait for human approval) that takes minutes, and only
then sends the follow-up that would have reused the (now large) conversation
prefix. Provider caches expire in minutes: Anthropic's default TTL is five
minutes~\cite{anthropic-caching}, OpenAI clears entries after ``5--10 minutes
of inactivity''~\cite{openai-caching}. The pause outlives the cache, and the
follow-up pays full price and full latency. At agent scale, dozens of such
pauses per task and prefixes of $10^5$ tokens, this is a real line item, and
trace analyses of production coding agents show the same pattern from the
serving side: long sessions repeatedly reuse large prefixes and put sustained
pressure on the KV cache that generic serving policies handle
poorly~\cite{cachewise}.

The defense is a \emph{keepalive}: during the pause, re-send the exact prefix
with minimal generation on a timer; each read refreshes the entry's TTL and its
recency in the eviction policy. The technique is established
practice~\cite{aider-keepalive,anthropic-prewarm}; what has been missing is a
quantified answer to the operator's actual questions. This paper provides it:

\begin{itemize}\itemsep2pt
\item \textbf{Does it work?} Yes, everywhere we measured. Across
\dataVendorsNice{}, prefix sizes of \dataSizes{} tokens, and idle gaps to
600\,s, a 30\,s keepalive held a median of ${\geq}\keepaliveMinHit\%$ of prompt
tokens cached in every provider-size-idle group; at 600\,s, where Anthropic's
baseline went 0/48 warm across three runs, the keepalive held 40/40
(\S\ref{sec:efficacy}).
\item \textbf{What frequency is economical?} Not the 30\,s convention.
Keepalive cost falls monotonically in the interval, so the optimum is the
largest interval safely under the TTL: ${\sim}4$\,min at Anthropic, an $8\times$
reduction in keepalive spend (\S\ref{sec:economics}).
\item \textbf{Which providers, and for how long a pause?} The keepalive saves
inside a \emph{paying band}: longer than the provider's eviction point, shorter
than $I_{\max} \approx \tau(w/r-1)$. We measure the eviction points (Anthropic
${\sim}5$\,min, DeepSeek ${\sim}10$\,min, OpenAI ${\sim}30$\,min, Google never)
and confirm the savings at a 30-minute pause: the keepalive saves on Anthropic
\emph{and} OpenAI, not Anthropic alone. DeepSeek's re-prefill is too cheap to
insure and Google's cache never reliably evicts, so for those two the keepalive
buys only latency (\S\ref{sec:economics}).
\item \textbf{What happens when everyone does it?} The incentive is individual
and immediate, so adoption is rational; but cache residency is priced per read,
not per token-hour, so a keepalive-saturated tier degrades for everyone. We
argue providers will be forced to meter residency (Google's explicit cache
already bills per token-hour), and we derive the builder's policy until then
(\S\ref{sec:equilibrium}).
\end{itemize}

\section{Background and provider parameters}
\label{sec:background}
Anthropic caches on an explicit \texttt{cache\_control} breakpoint: writes bill
at $1.25\times$ input, reads at $0.1\times$, default TTL 5 minutes with a paid
1-hour tier at $2\times$ write~\cite{anthropic-caching}. OpenAI caches
automatically above ${\sim}1024$ tokens at $0.1\times$ cached-read price, clears
after 5--10 minutes idle (longer off-peak), routes by prefix hash, and offers
\texttt{prompt\_cache\_key} for routing affinity~\cite{openai-caching}. Google
caches implicitly above ${\sim}2048$ tokens at ${\sim}0.25\times$ cached-read
price~\cite{google-caching}. DeepSeek caches automatically at $0.1\times$ read
price; its first-party cache is documented as disk-backed. In all four,
\emph{reading a cached prefix refreshes it}. Table~\ref{tab:params} collects
the parameters that drive the economics; the keepalive's whole mechanism is the
refresh-on-read semantics in the last row.

\begin{table*}[t]
\centering\footnotesize
\caption{Parameters driving keepalive economics (list prices, July 2026):
cached-read ratio $r$ and re-prefill ratio $w$ relative to input price,
documented idle TTL $T$, and the derived economical interval $\tau^{\ast}$ and
break-even horizon $I_{\max}$ (\S\ref{sec:economics}). Google's higher $r$
shrinks its break-even horizon.}
\label{tab:params}
\begin{tabular}{lcccccc}
\toprule
Provider & $r$ & $w$ & $T$ & $\tau^{\ast}$ & $I_{\max}$ & rent/hour at $\tau^{\ast}$ (100k prefix) \\
\midrule
Anthropic (5-min tier) & 0.10 & 1.25 & 300\,s & ${\sim}240$\,s & ${\approx}46$\,min & \$0.45 \\
Anthropic (1-hour tier) & 0.10 & 2.00 & 3600\,s & ${\sim}50$\,min & ${\approx}3.3$\,h & \$0.04 \\
OpenAI & 0.10 & 1.00 & 300--600\,s & ${\sim}240$\,s & ${\approx}36$\,min & \$0.19 \\
Google (implicit) & 0.25 & 1.00 & minutes & ${\sim}240$\,s & ${\approx}12$\,min & \$0.47 \\
DeepSeek & 0.10 & 1.00 & ${\leq}600$\,s & ${\sim}240$\,s & ${\approx}36$\,min & \$0.04 \\
\bottomrule
\end{tabular}
\end{table*}

\paragraph{The keepalive in production.}
We implement the defense in the \texttt{pi} agent harness as a per-tool-call
keepalive: pings run only while a tool batch executes and stop when it
completes (default off). The bounded form is the economically correct form
(\S\ref{sec:economics}).

\section{Related work}
Keepalive-style cache management is shipping practice: Aider added a cache
keepalive in 2024~\cite{aider-keepalive}, and Anthropic's documentation
recommends periodic pre-warming~\cite{anthropic-prewarm}. Practitioners have
also independently derived pieces of the economics we quantify: community cost
analyses of Anthropic's 5-minute TTL~\cite{brandonwie-ttl}, a documented
4-minute ping interval with a single-scenario break-even
estimate~\cite{veritas-coffee}, and keepalive plugins and proposals with
break-even arithmetic~\cite{cc-cache-keepalive,openclaw-keepalive}. These are
single-provider
(Anthropic), single-scenario, and computed rather than measured. On the
academic side, an evaluation of cache \emph{placement} for agentic
tasks~\cite{dontbreakcache} studies what to cache rather than how long it
lives. On the systems side, prefix-cache eviction is documented in vLLM's
design (LRU over content-addressed blocks)~\cite{vllm-prefix-caching}, studied
for batched inference in BatchLLM~\cite{batchllm}, characterized at
production scale in \emph{KVCache Cache in the Wild}~\cite{kvcache-wild}, and
studied for coding agents specifically in CacheWise~\cite{cachewise}, whose
trace analysis motivates reuse-aware eviction on the serving side. Our
question is the client-side complement of theirs: given caches we cannot see,
what can the agent itself do?
CacheProbe studies cache isolation across OpenRouter accounts~\cite{cacheprobe};
the multi-tenant billing externality of \S\ref{sec:equilibrium} was described
informally as ``prompt cache thrashing''~\cite{cache-thrashing-blog}. What is
missing is the measured, cross-provider account: retention and keepalive
efficacy across four providers with timing-gated data, whole-strategy costs
including the pings, and a per-provider policy. That is our contribution.

\section{Method}
\label{sec:method}
\paragraph{Harness.}
For each provider we send, per measurement \emph{cell}: reqA, a large prefix
with a unique per-cell salt (writing the cache); an idle gap of $I$ seconds, in
the keepalive condition replaying the identical prefix every 30\,s; and reqB,
the identical prefix, streamed to capture time-to-first-token. Anthropic,
OpenAI, and Google are queried first-party (Anthropic Messages API with an
explicit ephemeral breakpoint; OpenAI Chat Completions with
\texttt{prompt\_cache\_key} set to the cell salt; Gemini with implicit
caching). DeepSeek is queried via OpenRouter~\cite{openrouter,openrouter-caching}
pinned to a single backend
(DeepInfra); the pin is endpoint-level, so the served backend is recorded on
every call and any cell whose backend changes mid-cell is discarded.

\paragraph{Timing integrity.}
Every call records queue-entry, start, and completion timestamps, and the
quantity analyzed is the \emph{true idle} experienced by the cache entry
(\texttt{reqB.start}${}-{}$\texttt{reqA.end}), not the nominal schedule. Cells
launch staggered in bounded shuffled batches so offered load stays well under
the HTTP concurrency limit; a cell is valid only if reqB's local queue wait is
${\leq}5$\,s and the true idle is within $10$\,s of nominal. Every run contains
idle${}=0$ warm-reference cells, an immediate re-read of a just-written prefix,
and a run whose warm references miss is flagged and its timing re-verified
before inclusion. Baseline and keepalive cells run in \emph{separate time blocks}
(order alternated across replicates) so keepalive traffic cannot pressure the
tier during baseline measurement. In the runs reported here, median queue wait
was 0\,ms and median idle slip ${<}1$\,s across the valid cells.

\paragraph{Cost accounting.}
Every call (reqA, every keepalive ping, reqB) is priced and recorded.
First-party responses carry no cost field, so cost is computed from usage and
public list prices; DeepSeek uses OpenRouter's reported cost. Strategy cost is
reported both per-request (reqB) and whole-cell (reqA${}+{}$pings${}+{}$reqB).

\paragraph{Matrix and statistics.}
We test \dataVendorsNice{} at \dataSizes{} tokens, idle gaps of
\{\dataIdles\}\,s, $n{=}8$ samples per cell per run, in three independent
runs separated by 30 minutes to hours. Two follow-up experiments extend the
matrix: a \emph{retention sweep} (baseline-only, no keepalive, 100k prefix,
idle gaps of $360$--$540$\,s bisecting Anthropic's cliff and $900$--$2400$\,s
locating the later eviction points, $n{=}6$ per cell) and a \emph{paying-band
run} (baseline vs.\ $\tau^{\ast}{=}240$\,s keepalive at an $1800$\,s idle,
$n{=}8$, Google excluded since the sweep shows it has no eviction point). Both
use the same transports, validity gates, and warm-reference checks as the
matrix. The data is bimodal (a prefix is essentially warm
or evicted), so we report the \emph{warm rate}: the fraction of valid samples
whose reqB cached ${\geq}90\%$ of prompt tokens. Samples within a run share one
moment's tier conditions, so the run is the unit of independence: per-run
Fisher exact tests (Mann--Whitney as a continuous check), Bonferroni-corrected,
gated on a ${\geq}10$ percentage-point effect, and a claim is credited only if
it holds in every run.

\section{Results}

\subsection{The keepalive works}
\label{sec:efficacy}
\begin{figure*}[t]
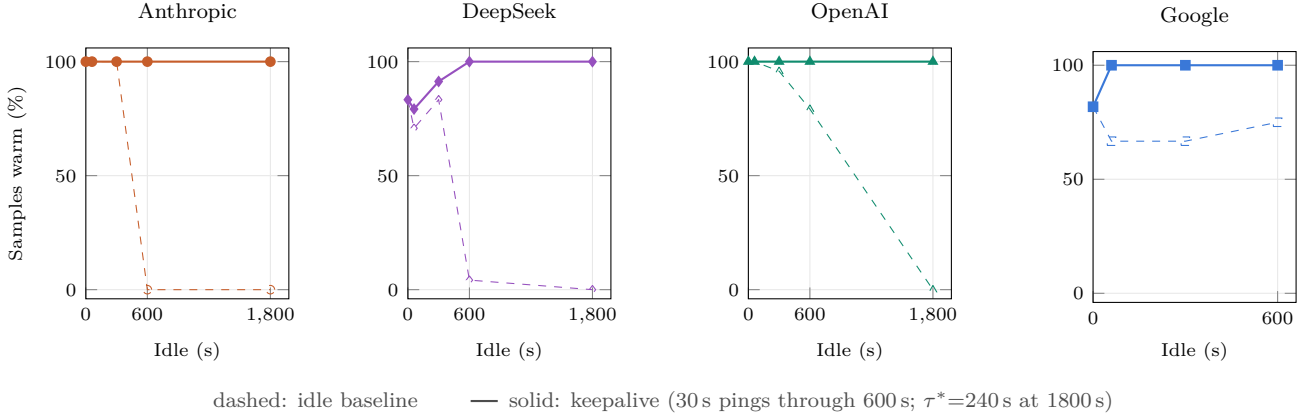

\centering
\newcommand{\warmpanel}[3]{%
\begin{subfigure}{0.24\textwidth}\centering
\begin{tikzpicture}
\begin{axis}[
  width=\linewidth, height=4.9cm,
  xlabel={Idle (s)}, xmin=0, xtick={0,600,1800},
  x tick label style={font=\scriptsize},
  ymin=-4, ymax=106, ytick={0,50,100},
  grid=both, grid style={gray!18}, tick label style={font=\scriptsize},
  label style={font=\scriptsize}, unbounded coords=jump,
  title style={font=\footnotesize}, title={#2}, #1]
\input{#3}
\end{axis}
\end{tikzpicture}
\end{subfigure}}
\begin{subfigure}{0.24\textwidth}\centering
\begin{tikzpicture}
\begin{axis}[
  width=\linewidth, height=4.9cm,
  xlabel={Idle (s)}, xmin=0, xtick={0,600,1800},
  x tick label style={font=\scriptsize},
  ymin=-4, ymax=106, ytick={0,50,100},
  grid=both, grid style={gray!18}, tick label style={font=\scriptsize},
  label style={font=\scriptsize}, unbounded coords=jump,
  title style={font=\footnotesize}, title={Anthropic}, ylabel={Samples warm (\%)}]
\input{data/plot-warm-100000-anthropic.tex}
\end{axis}
\end{tikzpicture}
\end{subfigure}\hfill
\begin{subfigure}{0.24\textwidth}\centering
\begin{tikzpicture}
\begin{axis}[
  width=\linewidth, height=4.9cm,
  xlabel={Idle (s)}, xmin=0, xtick={0,600,1800},
  x tick label style={font=\scriptsize},
  ymin=-4, ymax=106, ytick={0,50,100},
  grid=both, grid style={gray!18}, tick label style={font=\scriptsize},
  label style={font=\scriptsize}, unbounded coords=jump,
  title style={font=\footnotesize}, title={DeepSeek}, ]
\input{data/plot-warm-100000-deepseek.tex}
\end{axis}
\end{tikzpicture}
\end{subfigure}\hfill
\begin{subfigure}{0.24\textwidth}\centering
\begin{tikzpicture}
\begin{axis}[
  width=\linewidth, height=4.9cm,
  xlabel={Idle (s)}, xmin=0, xtick={0,600,1800},
  x tick label style={font=\scriptsize},
  ymin=-4, ymax=106, ytick={0,50,100},
  grid=both, grid style={gray!18}, tick label style={font=\scriptsize},
  label style={font=\scriptsize}, unbounded coords=jump,
  title style={font=\footnotesize}, title={OpenAI}, ]
\input{data/plot-warm-100000-openai.tex}
\end{axis}
\end{tikzpicture}
\end{subfigure}\hfill
\begin{subfigure}{0.24\textwidth}\centering
\begin{tikzpicture}
\begin{axis}[
  width=\linewidth, height=4.9cm,
  xlabel={Idle (s)}, xmin=0, xtick={0,600,1800},
  x tick label style={font=\scriptsize},
  ymin=-4, ymax=106, ytick={0,50,100},
  grid=both, grid style={gray!18}, tick label style={font=\scriptsize},
  label style={font=\scriptsize}, unbounded coords=jump,
  title style={font=\footnotesize}, title={Google}, ]
\input{data/plot-warm-100000-google.tex}
\end{axis}
\end{tikzpicture}
\end{subfigure}
\\[2pt]
{\footnotesize \textcolor{black!70}{dashed: idle baseline \qquad \rule[0.5ex]{1.2em}{0.8pt}\, solid: keepalive ($30$\,s pings through $600$\,s; $\tau^{\ast}{=}240$\,s at $1800$\,s)}}
\caption{\textbf{Per-provider warm rate vs.\ idle} (fraction of valid samples
whose post-idle reqB cached ${\geq}90\%$ of prompt tokens; 100k prefix; 40k
shows the same pattern on Anthropic, DeepSeek, and OpenAI; the $1800$\,s points
come from the paying-band run of Table~\ref{tab:bands}). Anthropic's baseline
collapses at $600$\,s (hard TTL) and DeepSeek's by $600$\,s; OpenAI's survives
$600$\,s and is gone by $1800$\,s; Google's never converges
(Fig.~\ref{fig:retention} has the full baseline sweep, including the gaps
between $600$ and $1800$\,s). The keepalive holds every provider at every
measured gap. Google was excluded from the $1800$\,s strategy run (no eviction
point, so no band to test); its keepalive points are $n{=}3$--$6$
(quota-limited); other points are up to $n{=}24$ across three runs, $n{=}8$ at
$1800$\,s.}
\label{fig:hit}
\end{figure*}

\begin{table}[t]
\centering\footnotesize
\caption{Fraction of samples whose post-idle \texttt{reqB} was warm (cache hit $\geq 90\%$), baseline vs.\ keepalive. Only cells passing all measurement-integrity gates are counted; conditions were collected in separate time blocks.}
\label{tab:summary}
\begin{tabular}{llrrr}
\toprule
Provider & Size & Idle & Warm$_{base}$ & Warm$_{ka}$ \\
\midrule
Anthropic & 40k & 60 & 24/24 & 24/24 \\
Anthropic & 40k & 300 & 24/24 & 21/21 \\
Anthropic & 40k & 600 & 0/24 & 20/20 \\
Anthropic & 100k & 60 & 24/24 & 24/24 \\
Anthropic & 100k & 300 & 24/24 & 19/19 \\
Anthropic & 100k & 600 & 0/24 & 20/20 \\
DeepSeek & 40k & 60 & 20/24 & 20/24 \\
DeepSeek & 40k & 300 & 16/23 & 21/21 \\
DeepSeek & 40k & 600 & 3/24 & 21/21 \\
DeepSeek & 100k & 60 & 17/24 & 19/24 \\
DeepSeek & 100k & 300 & 20/24 & 21/23 \\
DeepSeek & 100k & 600 & 1/24 & 21/21 \\
Google & 40k & 60 & 12/12 & 6/7 \\
Google & 40k & 300 & 7/12 & 5/5 \\
Google & 40k & 600 & 11/12 & 6/6 \\
Google & 100k & 60 & 8/12 & 7/7 \\
Google & 100k & 300 & 8/12 & 5/5 \\
Google & 100k & 600 & 9/12 & 3/3 \\
OpenAI & 40k & 60 & 24/24 & 24/24 \\
OpenAI & 40k & 300 & 22/24 & 20/21 \\
OpenAI & 40k & 600 & 20/24 & 23/23 \\
OpenAI & 100k & 60 & 24/24 & 24/24 \\
OpenAI & 100k & 300 & 23/24 & 24/24 \\
OpenAI & 100k & 600 & 19/24 & 23/23 \\
\bottomrule
\end{tabular}
\end{table}

Figure~\ref{fig:hit} and Table~\ref{tab:summary} are the efficacy result, and
the four providers sort into four regimes. \emph{Hard TTL} (Anthropic): the
baseline is warm through 300\,s and evicted to 0/48 samples across the three
runs at 600\,s, while the keepalive holds 40/40 at the same gap; the separation
is Bonferroni-significant in every run at both prefix sizes. \emph{Soft, lossy
cache} (DeepSeek via DeepInfra): the baseline evicts (4/48 warm at 600\,s) and
the keepalive holds 42/42, but short-idle samples show scatter we attribute to
machine-level routing inside the pinned endpoint, including occasional
keepalive misses at 60\,s that first-party providers did not produce; an
endpoint pin is not a machine pin, and a ping can warm one machine while reqB
lands on another. \emph{Slow cache} (OpenAI): its baseline survives
600\,s in most samples (39/48); it does evict, but on a longer timescale than
this window shows (\S\ref{sec:retention}). \emph{Lottery cache} (Google): its warm rate never
converges to warm or cold, sitting at 33--83\% at every gap
(\S\ref{sec:retention}); with no affinity lever and a commit delay, a Gemini
hit is a routing probability rather than a retention state. Google's one clean
signal is diurnal: its 600\,s baseline was 12/16 in the peak-evening run and
8/8 in the off-peak night run. Across all providers, sizes, and idles the
keepalive's median cache hit is ${\geq}\keepaliveMinHit\%$, and the ping logs
show the schedule was actually kept (median ping drift ${<}0.2$\,s). Our strict
bar (Fisher exact per run, Bonferroni over the 24 comparisons, required in
every run) is met by Anthropic at both sizes and DeepSeek at 100k; DeepSeek at
40k shows the same pooled separation but falls one run short of the per-run bar.

\subsection{Where each cache actually dies}
\label{sec:retention}
\begin{figure}[t]
\centering
\begin{tikzpicture}
\begin{axis}[width=\linewidth, height=5.4cm, xlabel={Idle time (s)},
  ylabel={Baseline warm (\%)}, ymin=-4, ymax=106, xmin=0, xmax=2450,
  xtick={0,300,600,1200,1800,2400}, x tick label style={font=\scriptsize,rotate=45,anchor=east},
  ytick={0,50,100}, tick label style={font=\scriptsize}, label style={font=\footnotesize},
  grid=both, grid style={gray!18}, unbounded coords=jump,
  legend style={font=\tiny, at={(0.5,-0.42)}, anchor=north, legend columns=4}]
\input{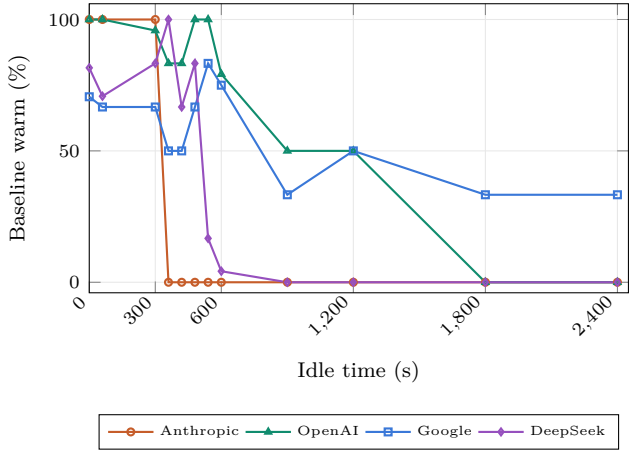}
\end{axis}
\end{tikzpicture}
\caption{\textbf{Retention curve: idle baseline warm rate out to 40 minutes}
(no keepalive, 100k prefix). No cache evicted before its documented TTL:
Anthropic falls off a cliff between $300$ and $360$\,s (its 5-minute TTL, no
grace period), DeepSeek collapses by $540$--$600$\,s, and OpenAI outlives its
documented 5--10 minutes, decaying gradually to fully cold by $1800$\,s.
Google never converges (a routing lottery, not a retention curve). A study
confined to gaps of $600$\,s or less would classify OpenAI as never-evicting;
its eviction point simply lies beyond that window.}
\label{fig:retention}
\end{figure}

The efficacy matrix measures whether the keepalive holds; the retention sweep
(Figure~\ref{fig:retention}) answers the prior question of when holding is even
needed: how long does an idle cache survive on its own? Sweeping baselines out
to $40$ minutes locates each provider's eviction point. Two facts fall out.
First, \emph{no cache evicted before its documented TTL, and none is
guaranteed past it}: Anthropic held $100\%$ at $300$\,s and was gone at
$360$\,s (its 5-minute TTL enforced with no grace period); DeepSeek held
through its window and collapsed by $10$ minutes; OpenAI outlived its
documented ``5--10 minutes'' (still ${\sim}50\%$ warm at $15$--$20$ minutes)
before going fully cold by $30$. The documented TTL is thus exactly what a
client may rely on and no more: warmth below it held everywhere we measured,
warmth above it is a provider-specific bonus that cannot be planned around.
The safe default without measuring is the documented TTL minus a margin. Second, OpenAI
is not the exception to the keepalive's value; it is a delayed instance of it.
Its cache does die, so past its eviction point there is a re-prefill to
insure against, and the keepalive can pay (\S\ref{sec:economics}).

\subsection{Keepalive economics: frequency and horizon}
\label{sec:economics}
Let $r$ be the cached-read price ratio, $w$ the re-prefill ratio ($1.25$ at
Anthropic, $1.0$ where re-caching is automatic and free), $T$ the TTL, and
$\tau$ the ping interval. Keeping a prefix alive through an idle $I$ costs
$(I/\tau + 1)\,r$ per input token; letting it die costs $w$ once. Everything
operators ask follows from this.

\paragraph{Frequency: ping as rarely as the TTL allows.}
Keepalive spend per unit time is $r/\tau$, strictly decreasing in $\tau$, and
the interval purchases nothing except TTL safety. The economical interval is
therefore $\tau^{\ast} = T - \mathrm{margin}$, where the margin covers ping
latency, jitter, and TTL-enforcement slack: ${\sim}4$\,min against Anthropic's
5-minute TTL (Table~\ref{tab:params}), an interval previously suggested in
practitioner work~\cite{veritas-coffee,openclaw-keepalive} and which we
validate experimentally below. The 30\,s convention in circulation
(including our own efficacy runs above) spends $8\times$ more than necessary
($7.8\times$ measured across our interval runs): holding a 100k-token Anthropic
prefix costs ${\sim}\$3.60$/hour at 30\,s pings versus ${\sim}\$0.45$/hour at
$\tau^{\ast}$. Where the TTL is fuzzy (OpenAI's
``5--10 minutes''), the uncertainty belongs in the margin, not the interval.

\paragraph{Horizon: the benefit has a ceiling.}
Break-even against re-prefill is $I_{\max} = \tau\,(w/r - 1)$:
${\approx}46$\,min for Anthropic's 5-minute tier at $\tau^{\ast}$,
${\approx}36$\,min for OpenAI and DeepSeek, and only ${\approx}12$\,min for
Google, whose implicit cache reads at $0.25\times$ rather than $0.1\times$
(Table~\ref{tab:params}). The keepalive saves inside a \emph{paying band}: an idle longer than the
provider's eviction point (or there is nothing to insure) but shorter than
$I_{\max}$ (or the pings exceed the re-prefill). Each provider's band follows
from its measured eviction point (\S\ref{sec:retention}) and its $I_{\max}$:
Anthropic ${\sim}6$--$46$\,min; OpenAI ${\sim}25$--$35$\,min against full
eviction, though its decay is gradual (half the samples are already cold at
$15$--$20$\,min, so in expectation the band opens earlier); DeepSeek
${\sim}10$--$36$\,min but with a re-prefill too cheap to be worth insuring; and
Google empty (no eviction point, so no band).

Two of our runs probe this band structure. \emph{At 600\,s}
(Table~\ref{tab:strategy}), only Anthropic and DeepSeek have evicted, so only
they have anything to insure; and even there the $30$\,s convention loses money
net of pings (its $I_{\max}$ at $30$\,s is only ${\approx}6$\,min), while the
$\tau^{\ast}{=}240$\,s arm cuts ping spend $7.8\times$ and saves $1.6\times$ on
Anthropic. A study confined to this gap would conclude the keepalive pays only
on Anthropic; that conclusion is an artifact of the gap, not a property of the
providers. \emph{At 1800\,s} (Table~\ref{tab:bands}), inside every evicting
provider's band, Anthropic, OpenAI, \emph{and} DeepSeek
are all fully cold at baseline, and the keepalive saves on both non-trivial
re-prefillers: $1.56\times$ on Anthropic at $\tau{=}240$\,s and up to
$2.45\times$ on OpenAI at its own optimal $\tau{=}480$\,s ($1.23\times$ at
$240$\,s). DeepSeek is the honest
exception: it evicts, but its cold re-prefill costs \$0.022, so seven pings cost
more than the eviction they prevent ($0.20\times$), a cost loss at every gap.
Its keepalive is a latency play, never a money one. The headline is therefore
not ``only Anthropic saves'' but ``every provider whose re-prefill is worth
insuring saves inside its measured band.'' Paid long-TTL tiers raise the ceiling
into the hours; no configuration makes indefinite warmth economical.

\IfFileExists{data/cost-table.tex}{\begin{table*}[t]
\centering\footnotesize
\caption{\textbf{The 10-minute pause: whole-strategy cost} (idle $600$\,s; reqA $+$ pings $+$ reqB, median, valid cells). At this gap only Anthropic and DeepSeek have evicted; OpenAI and Google are still warm at baseline, so their keepalive columns insure against nothing. The 30\,s convention loses money on every provider. Baseline/30\,s columns pool the matrix replicates ($n$ to 24); the 240\,s column is the interval-validation run ($n{=}4$).}
\label{tab:strategy}
\begin{tabular}{ll rrr rrr rrr}
\toprule
 & & \multicolumn{3}{c}{baseline} & \multicolumn{3}{c}{keepalive 30\,s} & \multicolumn{3}{c}{keepalive 240\,s} \\
\cmidrule(lr){3-5}\cmidrule(lr){6-8}\cmidrule(lr){9-11}
Provider & Size & cost & warm & TTFT & cost & warm & TTFT & cost & warm & TTFT \\
\midrule
Anthropic & 40k & \$0.267 & 0/24 & 1654\,ms & \$0.347 & 24/24 & 1240\,ms & \$0.166 & 4/4 & 1267\,ms \\
Anthropic & 100k & \$0.667 & 0/24 & 2285\,ms & \$0.867 & 24/24 & 1453\,ms & \$0.414 & 4/4 & 1332\,ms \\
DeepSeek & 40k & \$0.017 & 3/24 & 2332\,ms & \$0.100 & 25/25 & 1039\,ms & \$0.022 & 4/4 & 1388\,ms \\
DeepSeek & 100k & \$0.043 & 1/24 & 5398\,ms & \$0.249 & 25/25 & 1382\,ms & \$0.060 & 3/4 & 1950\,ms \\
Google & 40k & \$0.053 & 11/12 & 1846\,ms & \$0.291 & 10/10 & 1429\,ms & \$0.090 & 3/4 & 2117\,ms \\
Google & 100k & \$0.131 & 9/12 & 3952\,ms & \$0.649 & 7/7 & 1800\,ms & \$0.186 & 3/4 & 2561\,ms \\
OpenAI & 40k & \$0.046 & 20/24 & 1007\,ms & \$0.130 & 27/27 & 901\,ms & \$0.055 & 4/4 & 1074\,ms \\
OpenAI & 100k & \$0.115 & 19/24 & 1135\,ms & \$0.317 & 27/27 & 1051\,ms & \$0.136 & 4/4 & 1508\,ms \\
\bottomrule
\end{tabular}
\end{table*}
}{}
\IfFileExists{data/bands-table.tex}{\begin{table*}[t]
\centering\footnotesize
\caption{\textbf{The 30-minute pause: paying bands and the interval frontier} (idle $1800$\,s, 100k prefix; every evicting provider fully cold at baseline). Marginal cost of
the pause decision only: cold \texttt{reqB} if the cache dies vs.\
pings${}+{}$warm \texttt{reqB} if kept alive (sunk \texttt{reqA} excluded);
median, with warm counts. Cost falls as $\tau$ grows until the interval
crosses the provider's own retention; past that, each ping re-prefills a dead
cache at full price and the strategy costs \emph{more} than never pinging
(Anthropic at $\tau{\geq}480$\,s). $n{=}8$ at $\tau{=}240$\,s, $n{=}6$
otherwise.}
\label{tab:bands}
\begin{tabular}{lrrrrr}
\toprule
 & let die & \multicolumn{3}{c}{keepalive, marginal cost (warm)} & best saving \\
\cmidrule(lr){3-5}
Provider & cost & $\tau{=}240$\,s & $\tau{=}480$\,s & $\tau{=}900$\,s & (at $\tau$) \\
\midrule
Anthropic & \$0.333 & \$0.214 (8/8) & \$1.334 (0/6) & \$0.667 (0/6) & 1.56$\times$ (240\,s) \\
OpenAI & \$0.104 & \$0.085 (8/8) & \$0.042 (6/6) & \$0.162 (2/6) & 2.45$\times$ (480\,s) \\
DeepSeek & \$0.022 & \$0.108 (8/8) & \$0.070 (3/6) & \$0.043 (0/6) & none \\
\bottomrule
\end{tabular}
\end{table*}
}{}

\paragraph{The interval frontier: past the TTL, pings turn toxic.}
Since keepalive cost falls in $\tau$, why stop at $240$\,s? Because the
interval is bounded by the provider's \emph{own} retention, and
Table~\ref{tab:bands} shows what happens on either side of that line at the
30-minute pause. OpenAI's cache holds $480$\,s gaps, so $\tau{=}480$\,s halves
its keepalive cost again: \$0.042 marginal against a \$0.104 re-prefill, a
$2.4\times$ saving, double the $\tau^{\ast}{=}240$\,s figure. But its retention
does not support $\tau{=}900$\,s ($2/6$ warm), and Anthropic's 5-minute TTL
makes any $\tau{\geq}480$\,s actively destructive: each ping lands on a dead
cache and re-prefills at full write price, so the ``keepalive'' costs
$4\times$ more than never pinging at all (\$1.334 vs.\ \$0.333). An interval
past the TTL is not merely wasteful; every ping pays the full penalty it was
meant to avoid. The frontier confirms the prescription and sharpens it: the
economical interval is the provider's own measured retention minus a margin,
$\tau^{\ast}{\approx}240$\,s for Anthropic and DeepSeek but
$\tau^{\ast}{\approx}480$\,s for OpenAI, and a multi-provider agent must pick
$\tau$ per provider, not globally.

\paragraph{Latency: the unqualified win.}
Cost is the bounded benefit; latency is not. Warm reqB TTFT beats the cold
re-prefill on every provider where eviction occurs, and the gap widens at longer
idles: at 1800\,s (Table~\ref{tab:bands}) the keepalive cuts TTFT from
$2.9$ to $1.5$\,s on Anthropic, $4.4$ to $1.3$\,s on OpenAI, and $5.4$ to
$2.0$\,s on DeepSeek. For interactive agents this alone can justify the
keepalive, and it is the whole case for DeepSeek, whose cost band is empty.

\subsection{The operator's policy}
Keep the cache warm across pauses whose length is bounded and plausibly reused
(a tool call, an approval wait), at $\tau^{\ast}$ set \emph{per provider} to
its measured retention minus a margin (${\approx}240$\,s for Anthropic and
DeepSeek, ${\approx}480$\,s for OpenAI), and only inside the provider's paying
band: longer than its eviction point and shorter than $I_{\max}$. Below the
eviction point the baseline survives and the pings are waste; past $I_{\max}$
the pings exceed the re-prefill; and past the provider's retention the interval
itself turns toxic, each ping re-prefilling a dead cache at full price
(Table~\ref{tab:bands}). Where the re-prefill is too cheap to insure
(DeepSeek) or the cache never reliably evicts (Google), keep warm only for
latency, not cost. Use the paid long-TTL tier for hour-scale gaps; never keep a
dead session warm. This is the policy \texttt{pi} implements.

\section{Discussion: rational adoption, and the provider response it forces}
\label{sec:equilibrium}
We expect readers to implement keepalives, and they would be rational to: the
benefit is immediate, the cost is bounded by their own bill, and no provider
terms forbid it; providers in fact recommend pre-warming~\cite{anthropic-prewarm}.
But the equilibrium this points at is degraded, and the mechanism is worth
stating precisely. A cache tier's eviction policy ranks by expected reuse; a
keepalive manufactures recency, so once every client keeps its prefixes alive,
LRU has nothing left to rank and the tier degrades toward
first-in-first-out-of-luck. Residency today is priced per \emph{read}, not per
token held per second, and the price does not rise when the tier is hot. Each
operator's rational $r/\tau$ rent payment therefore imposes an unpriced
congestion cost on every other tenant: shorter effective TTLs, lower hit rates,
and the ``prompt cache thrashing'' spiral in which keepalive becomes mandatory
to extract any cache value at all~\cite{cache-thrashing-blog}. Individually
rational, collectively self-defeating.

Why do providers encourage the practice today? Because at current adoption the
exchange is paid and bounded: each refresh bills at the read rate (holding a
100k prefix alive at $\tau^{\ast}$ costs the client ${\sim}1.5\times$ the input
price per hour, recurring rent for the memory), the provider retains the right
to evict, and a cache-read ping converts a compute-expensive re-prefill into a
nearly free memory read, smoothing their compute demand. But the levers for the
congested regime are already visible, and we predict they will be pulled as
keepalive adoption spreads: absolute lifetime caps, per-account residency
quotas, paid long-TTL tiers (Anthropic's 1-hour tier at $2\times$ write is an
early step), and, the clean solution, metering residency per token-hour, which
Google's explicit context cache already does~\cite{google-caching}. Token-hour
pricing kills speculative warmth while leaving real reuse profitable: the rent
then tracks the resource actually consumed. The economic opportunity this paper
quantifies is therefore best understood as an arbitrage with an expiry date:
profitable now, and self-extinguishing at scale. Builders who adopt the bounded
policy of \S\ref{sec:economics} keep the benefit under every regime we can
foresee, including the metered one.

\section{Threats to validity}
\begin{itemize}\itemsep2pt
\item \textbf{Backend identity.} DeepSeek numbers describe the pinned
OpenRouter backend (DeepInfra), whose cache behavior is its own; an endpoint
pin is not a machine pin, which we observe as occasional keepalive misses at
short idles.
\item \textbf{Within-run correlation.} Samples in a run share one tier moment;
per-run $p$-values are descriptive, and only across-run replication is claimed.
\item \textbf{Google's quota.} Gemini cells were collected under a restrictive
account request quota (1K requests/day, which keepalive blocks repeatedly
exhausted): Google has two complete matrix runs (peak evening and off-peak
night) and one interval-validation run, plus partial cells from two
quota-killed runs; its keepalive arms remain thin ($n{=}3$--$7$ per cell). Its
implicit cache also offers no affinity lever, so warm rates carry
machine-lottery variance on top of the ordinary kind.
\item \textbf{Price drift.} Dollar figures use July-2026 list prices; ratios
and the form of the results are robust to drift.
\item \textbf{Load.} Retention timescales are measured at the load one
measurement client generates; provider tiers under fleet-wide keepalive
pressure are the subject of \S\ref{sec:equilibrium}, not of our retention
curves.
\end{itemize}

\section{Conclusion}
In agentic workloads the pause outlives the cache, and the keepalive, reading
the prefix back on a timer, is a proven, individually rational defense: it
holds prefixes warm across Anthropic, OpenAI, Google, and DeepSeek through gaps
that evict idle baselines, and it cuts the post-pause request cost by up
to \bestSavingRatio{}. The economical ping frequency is the largest interval
safely under the provider's own measured retention, about 4\,min at Anthropic
and 8 at OpenAI rather than the 30\,s convention, and past that retention the
interval turns toxic, each ping re-prefilling a dead cache at full price. The
saving has a hard ceiling at ${\approx}\tau(w/r-1)$, tens
of minutes at current prices. Because the benefit is real, universal adoption
is the rational equilibrium; and because residency is not metered, that
equilibrium degrades the shared tier until providers price cache residency
directly. One already does. The rest, we predict, will follow, and the sooner
operators understand the incentive, the sooner that day arrives.

\small
\bibliographystyle{plain}
\bibliography{references}

\end{document}